# Remarkable performance recovery in highly defective perovskite solar cells by photo-oxidation

Katelyn P. Goetz,[a] Fabian T. F. Thome,[c] Qingzhi An,[a] Yvonne J. Hofstetter,[a,b] Tim Schramm,[a,b] Aymen Yangui,[d] Alexander Kiligaridis,[d] Markus Loeffler,[e] Alexander D. Taylor,[a] Ivan G. Scheblykin,[d] Yana Vaynzof[a,b*]

a. *Chair for Emerging Electronic Technologies, Technical University of Dresden, Nöthnitzer Str. 61, 01187 Dresden, Germany*
b. *Leibniz-Institute for Solid State and Materials Research Dresden, Helmholtzstraße 20, 01069 Dresden, Germany*
c. *Kirchhoff Institute for Physics, University of Heidelberg, Heidelberg, Germany*
d. *Chemical Physics and NanoLund, Lund University, Lund, Sweden*
e. *Dresden Center for Nanoanalysis, Technical University of Dresden, Dresden, Germany*

*\*Prof. Dr. Yana Vaynzof, yana.vaynzof@tu-dresden.de*

Exposure to environmental factors is generally expected to cause degradation in perovskite films and solar cells. Herein, we show that films with certain defect profiles can display the opposite effect, healing upon exposure to oxygen under illumination. We tune the iodine content of methylammonium lead triiodide perovskite from understoichiometric to overstoichiometric and expose them to oxygen and light prior to the addition of the top layers of the device, thereby examining the defect dependence of their photooxidative response in the absence of storage-related chemical processes. The contrast between the photovoltaic properties of the cells with different defects is stark. Understoichiometric samples indeed degrade, demonstrating performance at 33% of their untreated counterparts, while stoichiometric samples maintain their performance levels. Surprisingly, overstoichiometric samples, which show low current density and strong reverse hysteresis when untreated, heal to maximum performance levels (the same as untreated, stoichiometric samples) upon the photooxidative treatment. A similar, albeit smaller-scale, effect is observed for triple cation and methylammonium-free compositions, demonstrating the general application of this treatment to state-of-the-art compositions. We examine the reasons behind this response by a suite of characterization techniques, finding that the performance changes coincide with microstructural decay at the crystal surface, reorientation of the bulk crystal structure for the understoichiometric cells, and a decrease in the iodine-to-lead ratio of all films. These results indicate that defect engineering is a powerful tool to manipulate the stability of perovskite solar cells.

**Introduction**

A commercial photovoltaic module should pay back the energy used to make it quickly and last for a long time thereafter. The high power conversion efficiencies (PCEs) of perovskite solar cells fulfill these requirements in part; however, their degradation under environmental operating conditions – where the solar module is exposed to oxygen, water, and temperature fluctuations – remains a critical challenge[1,2]. Researchers have engineered creative strategies to mitigate this problem, such as encapsulation layers which contain lead-chelating agents, simultaneously prolonging lifetime and preventing environmental contamination[3]. Nonetheless, a fundamental description of the degradation mechanisms of perovskite layers is incomplete, preventing their total protection.

Insight into this issue has been obtained by many avenues. Many past works examining the degradation of metal halide perovskites focused on the impact of humidity, which was identified early on to be a major source of instabilities[4–7]. Exposure to oxygen was not considered to be detrimental, especially considering that perovskite layers are often processed in dry air[8]; however, studies from Haque and colleagues reported that oxygen can diffuse through perovskite layers and lead to significant degradation by facilitating the conversion of $MAPbI_3$ to $PbI_2$[9–12]. Sun *et al.* demonstrated that this degradation is microstructure-dependent, as it commences at the grain boundaries of the perovskite layer[13]. Both studies suggested that such a degradation is mediated by an interaction with iodine vacancies in the perovskite lattice, which highlights the potential role of defects in determining the degradation pathway. A theoretical study by Meggiolaro *et al.* suggested that in iodine-rich perovskite compositions, exposure to oxygen may induce a healing effect of deep-level defects[14]. To experimentally link the properties of defects to the degradation of perovskite materials is challenging since their energetic and spatial location are difficult to detect; nonetheless, impurity clusters were recently observed to seed photo-chemical degradation in MA-free perovskites by multimodal imaging techniques[15]. An alternative approach is to intentionally introduce certain types of defects and explore how the degradation behavior changes. For example, we previously varied the defect profiles in $MAPbI_3$ perovskite films by fractionally tuning the stoichiometry of the precursor solution, thus inducing vacancies in understoichiometric samples and interstitials in overstoichiometric ones[16,17]. The impact of such changes on the stability was evident in their shelf storage behavior: understoichiometric solar cells were far longer-lived than overstoichiometric ones[16]. Such studies are not only indicative of the origin of environmental instability, but also allude to mitigation strategies.

Convoluting the experimental picture of degradation is that it is often, even necessarily, examined through the lens of the device. Not only does the perovskite degrade due to its own intrinsic chemical reactions to light, water, oxygen, and thermal stress, but so also do the extraction layers[18]. Furthermore, the perovskite may interact with the extraction layers during the degradation process. For example, it is well known that iodine can migrate through certain extraction layers and react with electrode materials like silver[19–21]. Here, it is not clear what proportion of the device performance has decayed due to increased parasitic recombination at the contacts, or due to degradation of the absorber layer itself, or due to other reasons.

In this work, we investigate the photooxidative response of methylammonium lead iodide perovskite in isolation of the electron extraction layers. We treat understoichiometric, stoichiometric, and overstoichiometric $MAPbI_3$ samples with oxygen and light and reveal a stark, defect-dependent contrast in perovskite behavior: samples which are understoichiometric exhibit degraded performance over time, samples which are stoichiometric show limited change in performance, and samples which are overstoichiometric show initially poor performance that recovers to maximum performance levels due to photooxidation. $MAPbI_3$ was used as a low-complexity composition to examine this effect and compare our observations to existing theoretical work; however, we find that similar healing behavior is observed for MA-free and triple cation perovskites, indicating this effect is general and applicable to state-of-the-art perovskite films. We use a combination of optical and structural analytical techniques to understand the reasons behind the observed trends, concluding that effective strategies to enhance the longevity of all perovskite compositions will protect and utilize iodine-related defects.

**Experiment**

Samples were made in the p-i-n (the so-called inverted) architecture, with the device structure shown in Fig. 1a. This is the same structure as we used in our previous work[16], with the exception of the hole-extraction layer. Here, the PEDOT:PSS was doped with additional PSS-Na to improve the open-circuit voltage[22]. The $MAPbI_3$ absorber layer was prepared using the lead-acetate trihydrate ($Pb(OAc)_2 \cdot 3H_2O$) recipe with added hypophosphorous acid (HPA)[23,24]. We have observed that the chemical lot of the $Pb(OAc)_2 \cdot 3H_2O$ precursor can impact the maximum PCE achieved for stoichiometric $MAPbI_3$ solar cells[25]; thus, our results here are based only on those precursors which regularly result in PCE's of 12% or more at an $MAI:Pb(OAc)_2 \cdot 3H_2O$ ratio of 3.0, with no hysteresis. Defects were introduced as previously[16]. In brief, starting with a precursor ratio of 2.96:1 of methylammonium iodide (MAI):$Pb(OAc)_2 \cdot 3H_2O$, two samples were spin-coated. A stock solution containing a known concentration of MAI was added to the precursor solution to raise its stoichiometry to 2.98 $MAI:Pb(OAc)_2 \cdot 3H_2O$, and the procedure was repeated until the stoichiometry reached 3.06. For each batch, we fabricated two samples of each stoichiometry of 2.96, 2.98, 3.00, 3.02, 3.04, and 3.06. Processing of the perovskite layer was accomplished in dry air (relative humidity < 0.5%). Extended fabrication details are included in the methods section. Note that although the MA here is added in large excess of 1 part per Pb atom, previous X-ray photoemission spectroscopy (XPS) and X-ray diffraction (XRD) measurements indicate that the excess MA is expelled during processing as MA(OAc), leading to a surface nitrogen content that is approximately the stoichiometric one part per lead atom[16]. Overstoichiometric samples do have a slightly higher N/Pb ratio at the surface than understoichiometric films.

To examine the degradation of MAPbI$_3$ with minimal impact from the electron transport layers, we fabricated the samples with only the substrate, indium tin oxide (ITO), doped PEDOT:PSS hole extraction layer, and perovskite absorber layer. Six samples – one of each specified stoichiometry – were placed in a sealed chamber with a gas mixture of 3% ± 0.5 % O$_2$ and 97% N$_2$. The chamber was placed under a solar simulator set to a light intensity of 1 Sun, adjusted against a NIST-calibrated silicon reference diode to the mismatch factor of the films (1.06-1.07). In parallel, six reference samples were placed in the dark in a nitrogen glovebox. All samples were finished together with the remaining layers ([6,6]-Phenyl C$_{61}$ butyric methyl ester, bathocuprine, and silver electrodes) after 10 hours in their respective environments and measured immediately. The third current-voltage (I-V) scan was used for measurement statistics, the effects of which are shown in Fig. S1 and S2. SI Note 1 further contains external quantum efficiency spectra (Fig. S3) and dark current density-voltage (J-V) (Fig. S4). Notably, the appearance of the samples exposed to oxygen and illumination showed no differences from the reference films visible to the naked eye. All films were brown in appearance.

## Results

### Degradation of Photovoltaic Cells

The photovoltaic figures of merit based on the forward voltage sweep for samples exposed to oxygen and illumination (red) and reference samples (black) are shown in Fig. 1a. Each point represents the average value for an entire substrate of up to 8 solar cells. As observed previously[16], the open-circuit voltage ($V_{OC}$) for the reference cells increases as a function of increasing stoichiometry. Only at a highly overstoichiometric MAI:Pb(OAc)$_2$ ratio (3.06) does this value decrease due to a high density of deep defects associated with iodine interstitials[17,26–28]. The treated cells show a similar increasing trend, but at a slightly lower average value. For the short-circuit current density ($J_{SC}$), there is a much stronger and stoichiometry-dependent response. For understoichiometric cells, the $J_{SC}$ drops to approximately half of the reference value of -21 mAcm$^{-2}$ upon photooxidation. Stochiometric samples display almost no degradation. On the other hand, overstoichiometric reference samples display a low $J_{SC}$, which is completely healed upon treatment with oxygen and light. The trend in the fill-factor (FF) shows only a small decrease for the under- and stoichiometric treated versus reference cells, while the overstoichiometric samples show improvement, like in the $J_{SC}$. Even more striking than the average values of the figures of merit are the J-V curves themselves, which are depicted on a representative basis in Fig. 1b. The overstoichiometric reference samples show high reverse hysteresis (summarized also in Fig. 1a), which is completely absent (cured) for the treated cells. This healing or degradation trend was not present for films aged for 10 hours in the dark under the gas flow of 3% O$_2$ and 97% N$_2$ , nor was it present for films aged under 100% N$_2$ flow and 1 Sun light exposure (Figs. S5-7), indicating that the response is due to the combination of oxygen and light. Note that N$_2$ exposure under light has previously been found to be beneficial as a post-deposition treatment, with elevated effects at temperatures of 80°C[29]. If such an effect is present here, it does not overcome the effect of the oxygen treatment.

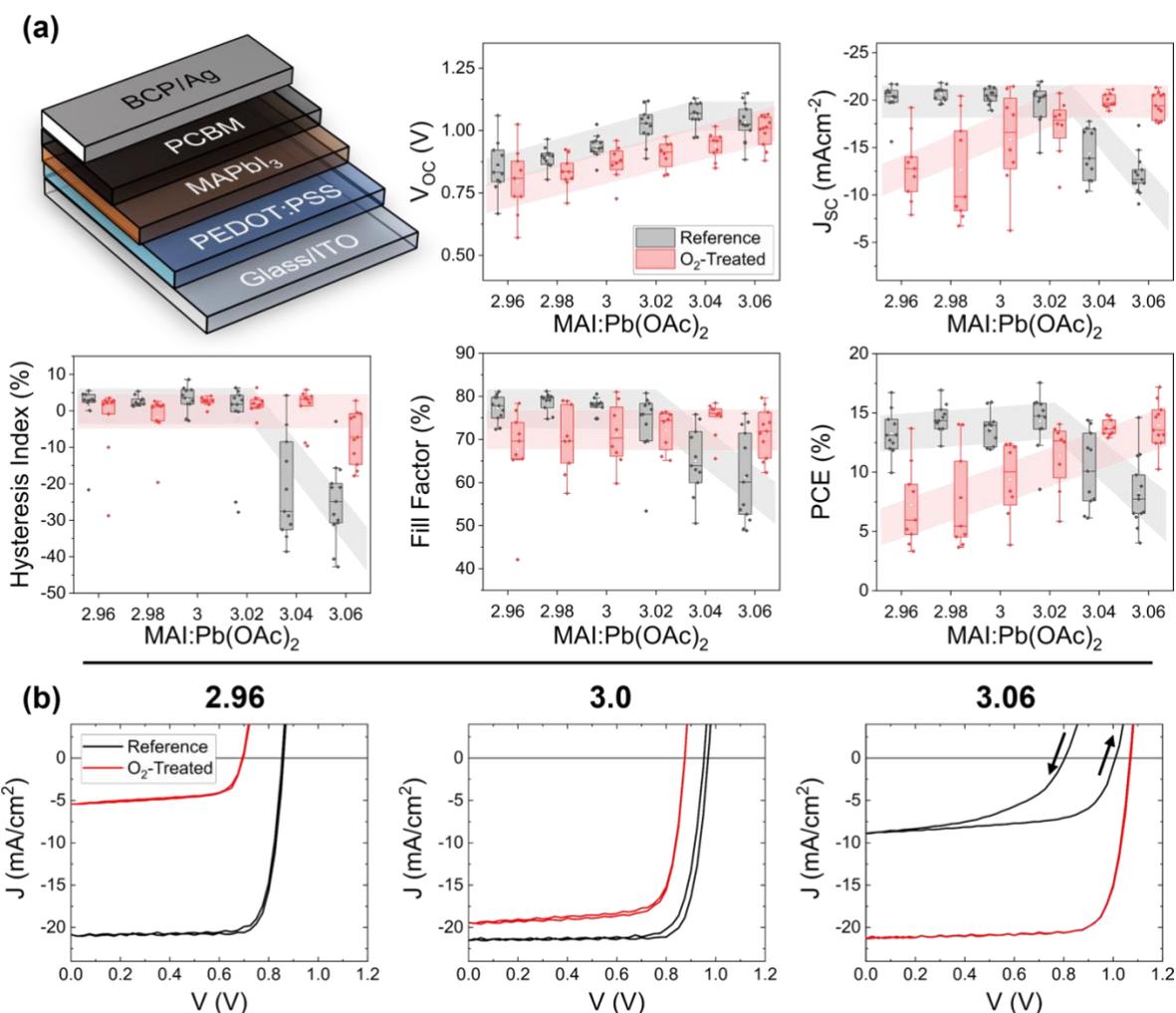

**Figure 1:** Photovoltaic performance summary based on the forward sweep of oxygen-treated perovskite films. The p-i-n device architecture is shown here from the top down. a) The VOC, JSC, FF, PCE, and HI are plotted for 11 untreated substrates (black) and 9 treated substrates (red). Each point represents the average figure of merit of up to 8 devices on a single substrate. The box gives the range of 25-75%, the vertical bars give 1.5 times the interquartile ratio, and the horizontal bar is the median. b) Representative JV curves for MAI:Pb(OAc)2 = 2.96, 3.0, and 3.06.

Although it does not currently yield record PCEs, $MAPbI_3$ presents a low-complexity composition, allowing us to better isolate the reasons for the observed changes. The lead acetate trihydrate recipe ensures a microstructure with compact crystalline grains spanning the entire thickness of the absorber layer, simplifying analysis. To assess the generality of the observed healing effect, we tested the photooxidative response of triple-cation recipe with excess of the iodine-containing precursors and in MA-free solar cells with excess formamidinium iodide (FAI). The PV characteristics are shown in Fig. S8-11 in the supplementary information. Both compositions show a lowered current density at higher stoichiometries, which is healed upon photooxidation. The triple cation cells show strong hysteresis that is healed upon oxygen exposure. This strongly suggests that iodine, rather than the methylammonium component, is active in the healing effect, and that it is applicable to all perovskite compositions. It is, however, notable that the effect of excess iodine on the PV characteristics is less extreme for both the triple-cation and the MA-free compositions versus $MAPbI_3$, and that a longer duration of oxygen and light exposure was required to observe healing in both compositions. This result is in agreement with studies that point to superior performance for mixed-halide perovskite compositions[30].

**Microstructure and Crystal Structure Changes**

To isolate and understand the physical and chemical processes leading to these effects, we examined the Glass/ITO/PEDOT:PSS/$MAPbI_{2.96-3.06}$ films by a variety of characterization techniques. Previous studies have implicated microstructural and crystallographic changes in the degradation of perovskite solar cells[13,24], as well as their recovery in the case of excess organic components[31,32]; therefore, we measured both treated and reference films using scanning electron microscopy (SEM). These images for representative understoichiometric, stoichiometric, and overstoichiometric samples are shown in Fig. 2. The grain size for the understoichiometric films is, on average, slightly lower than stoichiometric or overstoichiometric samples. We previously observed that small grains can decay faster than larger grains[13] but no signs of such decay are present here; therefore, we do not expect microstructure to account for the stoichiometry dependence of the photooxidation treatment. For films based on the higher ratio of MAI:$Pb(OAc)_2 \cdot 3H_2O$ in the precursor solution, darkened grain boundaries can be observed. This could be related to the accumulation of excess organic cation (MA) at the grain boundaries, as has been suggested previously[33,34]. For the films based on the 2.96 ratio, needles are frequently (but not always) seen on the film surface. These disappear at higher stoichiometries and could not be correlated with the degradation behavior of the solar cells. Comparing the reference cells to the degraded cells, the understoichiometric and stoichiometric cells undergo surface roughening upon exposure to oxygen and light. These results agree with previous observations of $MAPbI_3$ degradation under light and low-concentration oxygen flow[13]. The overstoichiometric cells, on the other hand, stay pristine, with isolated holes sometimes appearing throughout the films. No degradation of the bulk grains was found in cross-sectional imaging of multiple batches of films, as shown in Fig. S12 for films of 2.96, 3.02, and 3.06 precursor solution stoichiometry.

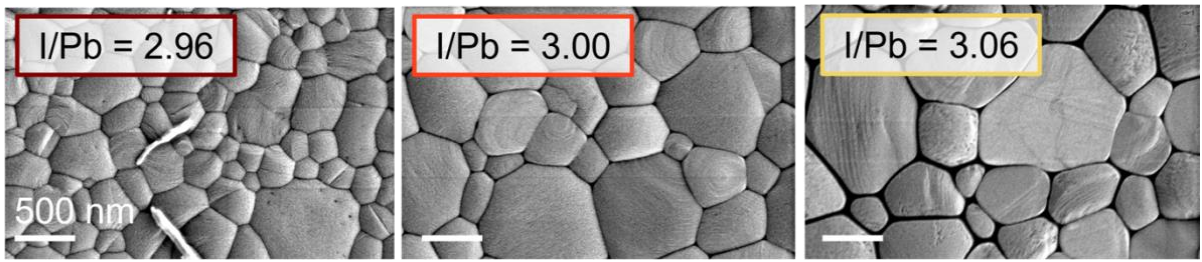
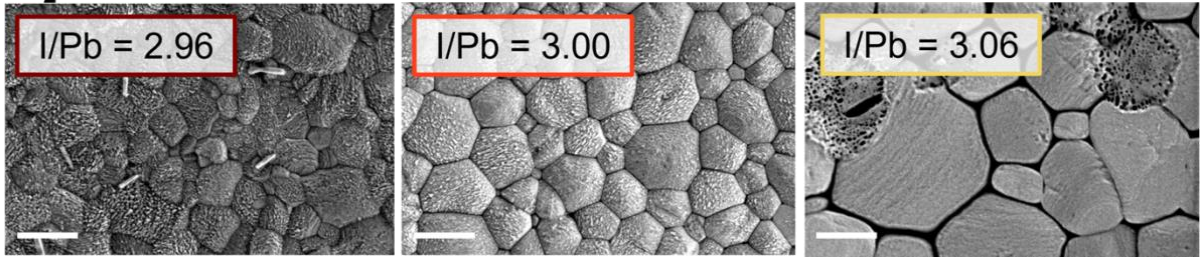

**Figure 2:** Representative SEM images of reference and oxygen-treated glass/ITO/PEDOT:PSS/Perovskite films for the understoichiometric, stoichiometric, and overstoichiometric cases.

Thin-film X-ray diffraction (XRD) measurements for reference and treated films of each stoichiometry are shown in Fig. 3. Peaks present at 14.1° and 28.5° match the 110 and 220 miller planes of the tetragonal $MAPbI_3$ single crystal structure[35]. The lack of other possible peaks indicates the preferential orientation of the films. The 2.96 stoichiometry does show some small peaks, which appear inconsistently from batch to batch. None of these could be assigned to the tetragonal phase or other low-dimensional phases of perovskite, or any suspected bulk precipitate defects, such as lead iodide (expected at 12.6°). It is possible but unproven that they are related to the needle-like crystallites appearing in the SEM. Interestingly, while the overstoichiometric samples retain their crystal structure upon photooxidation, the understoichiometric films show the emergence of peaks at 14.0° and 28.1°. This pattern corresponds to the 002 and 004 miller planes of the tetragonal perovskite single crystal structure, indicating a reorientation of the crystal structure of the films. A similar phenomenon has been observed for non-stoichiometric triple cation perovskites in the presence of moisture[32]. This was measured to be a surface effect via grazing-incidence wide-angle X-ray scattering (GIWAXs), whereas our measurements were completed in the θ-2θ configuration; thus, we suggest the reorientation observed here occurs in the bulk crystal.

Further insight into this process was gained by examining the degradation of the 2.96 films as a function of time (Fig. S13). Note that for this batch of films, the aforementioned peaks appearing occasionally at low intensity are more pronounced than those of Fig. 3. The short-circuit current density for one substrate of solar cells is shown in Fig. S13b to decrease steadily as a function of time. XRD changes (Fig. S13d) coincide with this result, showing a gradual emergence of the 002 or 004 peak. A shoulder becomes apparent at modest degradation time, 5 hours, and the intensity increases relative to the decreasing 110 or 220 peaks at longer times. These results agree with previous analyses showing that strain, misorientation, or otherwise non-uniform grain structure can reduce photocurrent[32,36,37]. Lead iodide is apparent at 15 hours for some samples at 12.6°. Meanwhile, the SEM (Fig. S13c) shows that the surface roughening begins almost immediately – it is already present at 2.5 hours degradation. Even at 15 hours, however, no bulk degradation could be observed in the cross-section images.

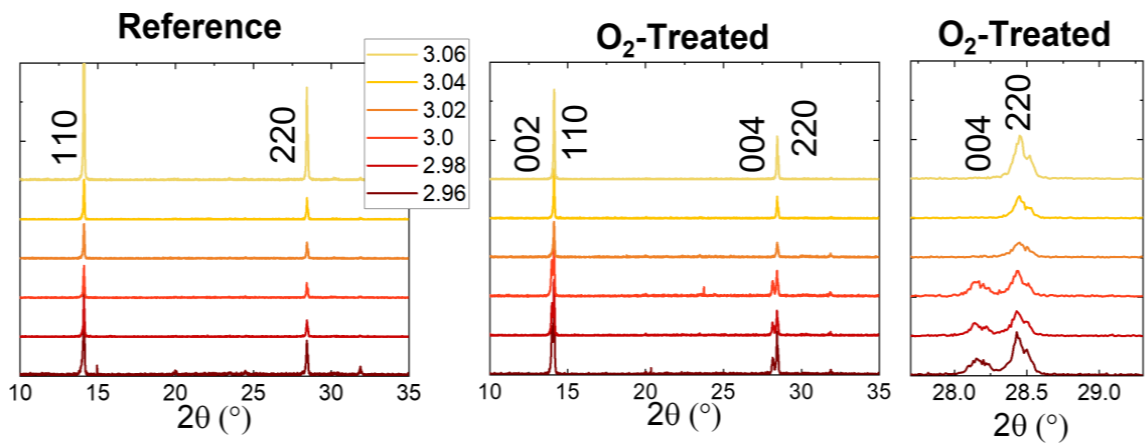

**Figure 3:** XRD for the reference and oxygen-treated samples, with an extra panel showing the emergence of the 004 peak near 28.5° in detail.

**Chemical Changes**

While enlightening for the understoichiometric samples, the SEM and XRD measurements give no direct insight into the dramatic improvement observed for the overstoichiometric samples. That the current density improves as holes appear throughout the microstructure (Fig. 2a) is even counterintuitive, suggesting a chemical or trap-related phenomenon. Therefore, we measured XPS in order to elucidate the chemical changes at the surface of the films. Fig. 4 shows the change in the ratios of the atomic percentages of Iodine, Nitrogen, and Oxygen to Lead, obtained by fitting the I 3d doublets, the N 1s singlet, the O 1s singlet, and the Pb $4f_{7/2}$ doublets, respectively, for an understochiometric (2.96) sample and an overstoichiometric (3.06) sample. In all cases, atomic percentages relative to Pb decrease upon oxygen and light treatment. No signs of PbO or $PbI_2$ are observed (Fig. S14). Notably, the quantity of iodine decreases by approximately the same amount for both under and overstoichiometric films. Although the proportion of nitrogen decreases less for the overstoichiometric case than the understoichiometric case, this suggests that the recovery of the overstoichiometric films occurs as their stoichiometry approaches the stoichiometric ratio of I/Pb. That nitrogen remains at roughly the same level coincides with the still-darkened grain boundaries observed by SEM in Fig. 2a, assuming that this is indeed due to MA accumulation and subsequent beam damage[34].

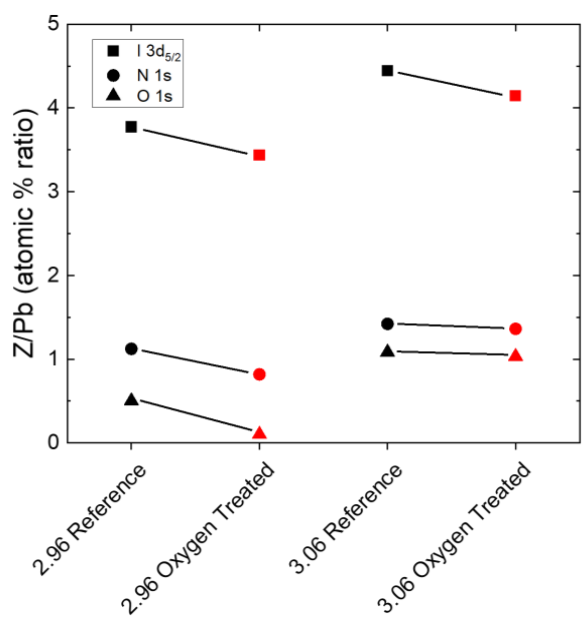

**Figure 4**: The atomic percentage ratios or iodine, nitrogen, and oxygen relative to the Pb4f$_{7/2}$ doublet for an understoichiometric and overstoichiometric sample. All ratios decrease upon photooxgen treatment.

**Trap-related Changes**

To understand the impact of the oxygen treatment on trap states, we investigated both steady-state photoluminescence (PL) and PL as dependent on pulse frequency and fluence. The PEDOT:PSS hole-extraction layer quenches PL; therefore, these experiments were performed on perovskite layers fabricated directly on glass. The grain size for these films increases relative to the films on Glass/ITO/PEDOT:PSS, meaning the degradation kinetics are likely different from those of the bulk cells; however, the relative behavior is still enlightening. The steady-state PL (532 nm) is shown in the supplementary information (Fig. S15) for all samples, both reference and those treated with oxygen and light. The reference samples show the same precursor dependence as previously observed, with understoichiometric reference films exhibiting the highest PL quantum efficiency (PLQE)[16,26]. The PLQE decreases with an increasing MAI:Pb(OAc)$_2$ ratio. Upon treatment, the PLQE increases relative to the untreated values for all films, with understoichiometric films still showing the highest absolute PLQE.

Expanding upon the steady-state measurement, we apply a recently introduced PLQE frequency and fluence mapping technique. This measurement gives a "fingerprint" of the sample in terms of the competition between the radiative recombination, charge trapping, trap-assisted non-radiative recombination, and the Auger processes[38]. PLQE maps (Figure 5) were obtained by scanning the repetition rate of the laser from 100 Hz to 80 MHz – almost 6 orders magnitude range – for 5 pulse fluences P1 (not shown), P2, P3, P4 and P5. The total difference between the fluence P2 and P5 is approximately three orders of magnitude. Overstoichiometric (3.06) reference perovskite films are shown as black squares, overstoichiometric (3.06) films treated with oxygen and light are shown as filled red triangles, and stoichiometric (3.00) reference films are shown as empty red squares. While the stochiometric sample map is very similar to what was previously measured[38], the overstoichiometric sample map is drastically different. We notice a strong decrease (up to one order of magnitude) of PLQE at low excitation power conditions (average power less than 1 Sun). This difference, however, becomes less pronounced upon increasing the excitation power. The PLQE of both the overstoichiometric and the stoichiometric reference samples becomes similar for high excitation power density (the right side of the plot). This behavior of the overstoichiometric sample is what would be expected for perovskite samples with a much higher defect concentration leading to a significantly stronger non-radiative recombination in comparison with the reference sample. These defects are likely to be iodine interstitials, which are known to lead to deep traps states[27,28] in $MAPbI_3$ and cause a reduction in PLQE[17,26]. However, these non-radiative channels are not able to compete with the radiative and Auger recombination at high fluences resulting in similar PLQE for both samples at high excitation power density.

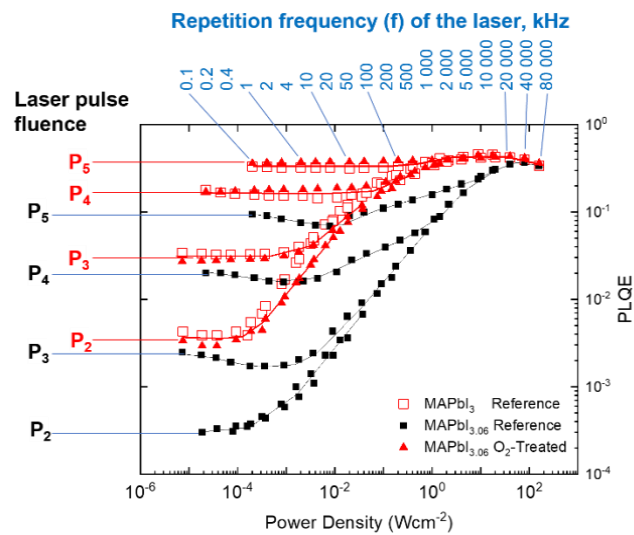

**Figure 5:** PLQE as a function of fluence and pulse frequency for reference samples of overstoichiometric (black squares), stoichiometric films (empty red squares), and verstoichiometric samples treated with oxygen and light (red triangles). The oxygen-treated overstoichiometric sample follows the same profile as the stochiometric reference (untreated) sample. Pulse fluences are $P_1=4.1 \times 10^8$ (not shown), $P_2=4.9 \times 10^9$, $P_3=5.1 \times 10^{10}$, $P_4=5.5 \times 10^{11}$ and $P_5=4.9 \times 10^{12}$ photons/cm$^2$.

It is remarkable that the overstoichiometric sample after exposure to oxygen and illumination shows practically the same PLQE map as the reference sample. This means that the oxygen and light treatment leads to the recovery of the overstoichiometric sample to quantitatively the same recombination regimes over the entire range of charge-carrier concentrations probed in the PLQE mapping experiment, as is observed for a stoichiometric perovskite film. This elimination of the additional recombination mechanisms explains why the PV performance of $O_2$-treated overstoichiometric solar cells is very similar to that of stoichiometric reference devices.

It should be noted that the trends of the PLQE oppose those of the $V_{OC}$ (Fig. 1). While the PLQE decreases as iodine content increases, the $V_{OC}$ increases; and while the $V_{OC}$ decreases due to light and oxygen treatment, the PLQE increases. The former effect has been observed before and attributed to the higher built-in potential of the more overstoichiometric films[16]. The higher iodine content creates more PLQE-lowering trap states, but it also causes better energetic alignment with the PCBM electron transport layer, which results in the higher $V_{OC}$. In the case of the oxygen treatment, XPS shows that the quantity of iodine at the surface decreases for both high and low stoichiometries, which we suggest corresponds to a lower density of deep iodine interstitial ($I^-$) trap states. The reduced iodine would thus, as previously, reduce the built-in potential and the $V_{OC}$ of the device while simultaneously decreasing non-radiative recombination and raising the PLQE. Additional PLQE enhancement for the understoichiometric and stoichiometric films could also arise from enhanced light-outcoupling due to the surface roughening (Fig.2).

**Discussion**

To summarize our experimental results, we observed that understoichiometric films exposed to light and oxygen degrade. Solar cells show reduced $J_{SC}$ and slightly reduced $V_{OC}$, which correlates with surface roughening and reorientation of the bulk crystal. Lead iodide is not detected by XPS or XRD on films exposed to oxygen under illumination for 10 hours, but is seen by XRD at 15 hours. Stoichiometric solar cells are stable on the timescale studied, showing some surface roughening. Very overstoichiometric solar cells show initially low $J_{SC}$ and strong hysteresis which becomes absent upon oxygen treatment. There is no strong change in the crystal structure, and the microstructure shows the surprising appearance of holes within the film. Both under and overstoichiometric samples show similar reduction in the I/Pb ratio upon exposure to light and oxygen, and all films show a relative improvement in their PLQE values.

These results can be brought together with existing literature on $MAPbI_3$ degradation to understand the role of defects. The initial experiment exploring the change in precursor stoichiometry for the lead acetate trihydrate recipe examined shelf storage of the solar cells – that is, the storage of fully complete cells in the dark under ambient conditions, measured over the course of several months[16]. The trends from shelf storage are the opposite of the present study. Overstoichiometric cells decayed fastest, while understoichiometric cells showed an in initial improvement that was followed by decay. This contrasting response is due to the presence of the electron withdrawing layers in the shelf-storage study. Aristidou and colleagues have examined the diffusion of oxygen into $MAPbI_3$ under light excitation, proposing a degradation mechanism where the oxygen diffuses into iodine vacancies, and, following the photoexcitation of the perovskite, accepts an electron to form the $O_2^-$ superoxide[9]. Following the deprotonation of the methylammonium, the perovskite decomposes into $H_2O$, $CH_3NH_2$, $PbI_2$, and $I_2$. The presence of electron withdrawing layers – $TiO_2$ or, in the present case, PCBM – mitigates this reaction for understoichiometric films despite the likely higher density of iodine vacancies[12].

In the case of the unprotected films in this experiment, it might seem likely that solid degradation products like $PbI_2$ should be observed if such a mechanism were at play. For the understoichiometric cells, the degradation of the photovoltaic characteristics – most pronounced in the form of $J_{SC}$ degradation – occurs before any appreciable changes to the lead chemistry detected by lab-scale XPS or XRD. Given that $PbI_2$ is seen shortly after the timescale of this experiment, at 15 hours (Fig. S13), it is probable that it is present at earlier times at lower levels. We therefore suggest that the key inhibitors for the passage of current are the surface roughening and reorientation of the crystal structure, both of which could cause resistive barriers to transport.

In the case of stoichiometric and overstoichiometric cells, the relatively low density of iodine vacancies would protect against the infiltration of oxygen discussed above for the case of the understoichiometric cells. Relevant to the response of the overstoichiometric cells to the oxygen treatment, the formation of $I_2$ in perovskite materials has also been reported as the result of the oxidation of $I^-$ in perovskite layers exposed to electrical bias, in the absence of elemental oxygen[39,40]. Researchers proposed that the $I_2$ in this case remained trapped in the film in the form of polyiodides and could be reversibly reduced[39]. Such a halide oxidation reaction was also found to seed halide redistribution in mixed-halide compositions[41]. The electrochemical potential for this reaction is very low – 0.35 V versus the normal hydrogen electrode (NHE) – making it favorable well below open-circuit conditions or even under light biasing. In our case, the reduction partner would be oxygen itself rather than a cathode under bias, and the $I_2$ and other decomposition products, like water, are likely eliminated under the carrier gas flow.

Taken together, we suggest the degradation of understoichiometric $MAPbI_3$ proceeds via oxygen infiltration of the iodine vacancies and subsequent degradation, as has been reported previously. In the case of overstoichiometric $MAPbI_3$, excess iodine in the form of iodide ($I^-$) interstitials is eliminated during the exposure to oxygen and illumination. We speculate that this occurs by an oxidation reaction to form volatile $I_2$, which leaves the surface of the perovskite layer under carrier gas flow.

These results have broad implications for a variety of perovskite applications. Because such a degradation mechanism is seen in devices under bias, our observations of defect-mediated degradation are likely relevant to even anoxic device fabrication and operation. The healing of highly overstoichiometric perovskites can result in more error-tolerant processing conditions. Importantly, it can be applied post-film fabrication and prior to the application of electrodes, making it compatible with multilayered device architectures. As suggested by the examination of triple-cation and MA-free films, its application to other perovskite or perovskite-additive compositions would need specific optimization.

## Conclusions

In conclusion, we demonstrate that perovskite films with overstoichiometric defect profiles can heal under exposure to environmental conditions. The behavior of the understoichiometric cells upon exposure to oxygen and light is what would be intuitively expected – the cells degrade steadily due to extended exposure to the environment. Only somewhat surprising is that the PCE diminishes before the lead iodide is detectable by XRD or XPS. Meanwhile, overstoichiometric cells improve drastically when exposed to oxygen under illumination. XPS suggests a global reduction in the I/Pb ratio, indicating that this ratio becomes more stoichiometric. Likewise, PLQE mapping suggests similar trap profiles and recombination processes for untreated stoichiometric cells and treated, overstoichiometric cells, indicating that a reduction in a deep trap state, such as iodine interstitials, is responsible for the healing. Our results suggests that defect management and engineering can be a useful strategy toward fabricating stable, high-efficiency perovskite solar cells.

## Experimental Methods

### Sample Fabrication Methodology

**$MAPbI_3$ Sample Preparation.** Glass substrates with pre-patterned indium tin oxide (ITO, PsiOTech Ltd., 15 Ω Sq$^{-1}$) were cleaned by sequential sonication in soap water, de-ionized water, acetone, and isopropanol, for 10 min each. The samples were rinsed with isopropanol, dried with nitrogen, and inspected to ensure a lack of visible residue. They were then treated for 10 minutes with oxygen plasma.

The hole-extraction layer was prepared immediately following the substrate cleaning. Poly(sodium-4-styrene sulfonate) (PSS-Na, Sigma Aldrich) was dissolved in de-ionized water to a concentration of 15 mg mL$^{-1}$. An aqueous suspension of PEDOT:PSS (poly(3,4-ethylenedioxythiophene) polystyrene sulfonate) (*Clevios* P VP AI 4083, Heraeus) was added to the PSS-Na solution in a volume:volume ratio of 4:1 for PSS-Na solution:PEDOT:PSS stock suspension. This was spin-coated in air onto the cleaned substrates at 4000 RPM for 30 s, and annealed at 150 °C for 15 min.

The $MAPbI_3$ layer was prepared by the method established by Fassl *et al.*[16] The perovskite precursor solution was prepared by weighing lead acetate trihydrate ($Pb(OAc)_2 \cdot 3H_2O$, Sigma Aldrich or TCI) and methylammonium iodide (MAI, GreatCell Solar) in a ratio of 2.96:1 MAI:$Pb(OAc)_2 \cdot 3H_2O$ and dissolving in *N,N*-Dimethylformamide (DMF, anhydrous, Sigma Aldrich) to a 42 wt % (weight percent) concentration. The precursors were weighed in air and the DMF was added in a dry-air box (relative humidity < 0.5%). Hypophosphorous acid (Alfa Aesar) was added to the perovskite solution to the amount of 1.7 μL per 100 mg of MAI precursor, also in the dry-air box. To maintain the same impact of weighing and pipetting error from batch to batch, all batches were based on an MAI weight of close to 200 mg. In parallel, a stock solution containing MAI dissolved in DMF was prepared at a concentration of 29% by weight. To prepare samples of incrementally varying precursor solution stoichiometry, two were spin-coated from the 2.96 MAI:$Pb(OAc)_2 \cdot 3H_2O$ solution. A known quantity of the MAI stock solution was then added to raise the stoichiometry to 2.98, and two more samples were spin coated. The process of adding stock solution and spin coating was repeated until a stoichiometry of 3.06 was reached, for a total of 12 samples.

To prepare the perovskite films, the precursor solution was spin-coated onto the hole-transport layer at 2000 RPM for 60 s in a dry-air box. During this process, the spin-coater chamber was kept closed, with the hole at the top of the lid blocked. The sample was blow-dried with compressed dry air for 20 s immediately after the spin-coater braked, and then rested at room temperature for 5 min. The films were then annealed at 100 °C for 5 min.

Following either the oxygen and light treatment or storage in a dark, nitrogen atmosphere, the remaining layers were applied to all 12 solar cell samples. In a nitrogen glovebox, PCBM ([6,6]-Phenyl $C_{61}$ butyric acid methyl ester, Solenne) was dissolved in chlorobenzene (anhydrous, Sigma Aldrich, 20 mg/mL). This was dynamically spin-coated onto a substrate at 2000 RPM for 30 s. The films were annealed at 100 °C. After these were cooled, a hole-blocking layer of bathocuprine (BCP, Sigma Aldrich, 0.5 mg/mL in anhydrous isopropanol) was dynamically spin-coated at 4000 RPM for 30 s. Silver electrodes were evaporated to a thickness of 80 nm at a rate of 0.1-0.5 Å/s. The final electrode area was 1.5 mm x 3 mm.

Samples for photoluminescence measurements were fabricated on glass. Microscope slides were cut into approximately 1 $cm^2$ and cleaned by the same method as the ITO/glass substrates. The perovskite layer was prepared in the same way as the solar cells.

**Triple Cation and MA-Free Perovskite Sample Preparation**. To test the composition-dependence of the photooxidative response, we fabricated triple-cation perovskite solar cells and MA-free perovskite solar cells. Glass/ITO substrates were cleaned as in section SI1. PTAA (poly(triaryl amine), Sigma Aldrich) was spin coated in a dry-air box from a 1.5 mg/mL solution in anhydrous toluene at 2000 RPM for 30 s, and subsequently annealed at 100 °C for 10 min. 30 µL of anhydrous DMF was spin coated onto the PTAA at 3000 RPM for 30 s to achieve appropriate surface wetting for the perovskite layer.

The triple-cation precursor solution was prepared as previously based on the stoichiometric compound $Cs_{0.05}(FA_{0.83}MA_{0.17})_{0.95}Pb(I_{0.9}Br_{0.1})_3$[42]. PbI$_2$ and PbBr$_2$ (TCI) were dissolved in a mixture of DMF:DMSO (4:1 v:v, where DMSO is dimethyl sulfoxide). CsI (TCI) was dissolved in pure DMSO at 125 °C. As achieving the precise stoichiometric ratios is paramount, the concentration of all solutions must be highly accurate. To account for the expansion of the solution upon precursor dissolution, known volumes of solutions were previously weighed, from which the true concentration calculated. This procedure used an initial concentration of 1.26 M. For additional information regarding this process, see reference 3[43]. These solutions were then mixed in a volume ratio of CsI:PbI$_2$:PbBr$_2$, 0.05:0.85:0.15 resulting in a 1.2 M solution of $Cs_{0.05}Pb(I_{1.75}Br_{0.3})$. This forms a so-called inorganic stock solution. MAI and FAI (GreatCell Solar) were then weighed in separate vials. To achieve a stoichiometric sample, the inorganic $Cs_{0.05}Pb(I_{1.75}Br_{0.3})$ stock solution was added to the MAI and FAI-containing vials to achieve a molar ratio of 0.95:1 of inorganic components to organic

(MAI and FAI). These FAI-inorganic and MAI-inorganic solutions were then mixed at a volume:volume ratio of 5:1 FAI-inorganic:MAI-inorganic to achieve the final perovskite precursor solution.

Stoichiometry was tuned by adjusting the ratio of [MAI + FAI]:Pb in the precursor solutions. Initially, a smaller than stoichiometric amount of inorganic stock solution was added to the weighed MAI and FAI to achieve the condition of overstoichiometric. Following the spin-coating of initial samples, the overstoichiometric solution was diluted with known quantities of the inorganic $Cs_{0.05}Pb(I_{1.75}Br_{0.3})$ solution, similar to the $MAPbI_3$ samples. Here, x was varied in the formula $Cs_{0.05}(Fa_{0.83}Ma_{0.17})_{0.95 \cdot x}Pb(Br_{0.1}I_{0.9+(x-1) \cdot 0.95/3})_3$ from the overstoichiometric 1.12 to 1.06.

Films were fabricated by spin coating 30 µL of the perovskite solution of known stoichiometry at 1000 RPM for 10 s, increased to 5000 RPM for 30 s. 10 s before the end of the 5000 RPM step, 200 µL of the anti-solvent TFT (α,α,α-Trifluorotoluene) was quickly dripped onto the substrate. Devices based on this antisolvent are dripping-speed independent[42].

The MA-free perovskite solution was prepared in a similar fashion to the triple-cation perovskite solution, omitting the MAI. The 1.1 M solution of $Cs_{0.1}FA_xPb_{2+x}I_{2.9}Br_{0.1}$ was prepared at an initial understoichiometric x = 0.87 and sequentially spin coated and diluted with known quantities of a stock solution of FAI in DMF:DMSO (4:1 v:v). Spin-coated occurred by the same method as the triple cation cells, using the extrusion-speed-independent antisolvent chlorobenzene[44]. Cells were finished after light and oxygen treatment with PCBM, BCP, and silver electrodes in the same fashion as the $MAPbI_3$ cells.

**Light and Oxygen treatment.** For the oxygen and light treatment, samples were placed in a stainless-steel chamber with a transparent window at the top. A constant flow of 97% nitrogen and 3% oxygen, ±0.5%, was connected to the chamber. The oxygen and water content of the atmosphere were continuously monitored using a zirconia sensor (Cambridge Sensotet, Rapidox 2100). The samples were exposed to 100 mW/cm$^2$ simulated AM1.5 sunlight (Abet Sun 3000, class AAA), adjusted to account for the window and the mismatch factor of the perovskite films (1.06-1.07, i.e., the exposure conditions were the same as the measurement conditions).

For the triple cation and MA-free compositions, the duration of the treatment was 20 h. Films proved more resilient to the treatment than the $MAPbI_3$ films, possibly due to the addition of Br.

**Solar Cell Characterization**

Current-voltage characteristics were measured for all solar cells consecutively, immediately following the final fabrication steps. A Keithley source-measure unit was used to apply voltage and measure current under AM 1.5 simulated sunlight with 100 mW/cm2 irradiation (Abet Sun 3000, class AAA). The light intensity was calibrated with a Si reference cell (NIST traceable, VLSI) and corrected using the measured spectral mismatch factor of 1.06-1.07. The cells were scanned from forward bias to short circuit and back (1.2 V- 0V) with a step size of 0.025 V and a dwell time of 0.1 s at each step. The cells were measured three times consecutively; values from the third measurement are plotted in the main text, Fig. 1.

**Scanning Electron Microscopy**

Scanning electron microscopy was measured using a Gemini 500 (Zeiss) at an accelerating voltage of 1.5 kV.

**X-Ray Diffraction**

Perovskite films on Glass/ITO/PEDOT:PSS were measured in the θ-2θ configuration using a Bruker D8 Discover diffractometer. The X-ray source was $Cu_{k\alpha}$.

**X-Ray Photoelectron Spectroscopy**

X-ray photoelectron spectroscopy spectra were acquired by a PHOIBOS 100 analyzer system (Specs, Berlin, Germany) at a base pressure of $10^{-11}$ mbar using an Al Kα excitation (1486.6 eV).

**Photoluminescence Spectroscopy**

**Steady-state measurements.** Steady-state photoluminescence quantum efficiencies (PLQEs) of perovskite films on glass were evaluated using the method by de Mello, et al[45]. We used a 532 nm laser at 10 mW power. Perovskite films for this measurement were fabricated directly on glass and measured under nitrogen for 15 minutes. The same films were placed in the degradation apparatus and treated for 15 hours. These were measured again under the same PL measurement conditions.

**Photoluminescence Mapping.** The photoluminescence frequency and fluence maps of the perovskite films were measured in a custom-built photoluminescence microscopy setup, see the details of the method elsewhere[38]. The samples were placed under N2 gas flow and were excited through an objective lens (Olympus 40X, NA = 0.6) using a 485 nm pulsed laser (Pico Quant, 150 ps pulse width) which the excited area of 30 micrometers in diameter. The emitted light was collected the same objective and detected by a CCD detector. Each PLQE(f,P) map included 85 combinations of laser frequencies and pulse fluences. Five different pulse fluences were used, namely P1=4.1x10^8, P2=4.9x10^9, P3=5.1x10^10, P4=5.5x10^11 and P5=4.9x10^12 photons/cm2, while the repetition frequency of the laser (f) was varied between 100Hz and 80MHz.

    To measure a full PLQE(f,P) map, the CCD acquires a series of PL images, one for each combination of the laser repetition rate f and the pulse fluence P. The full measurement was done in an automatic regime using a LabVIEW software which was reading and implementing the pre-prepared list of parameters designed for each individual measurements (repetition rate, excitation and emission filters, exposure time of the camera, etc.). The exposure time per data points was 0.1 s for the high excitation power and hundreds of seconds for low excitation power. The total time required to measure one map was about 2-3 h depending on the parameters.

## Author Contributions

YV designed and supervised the project. KPG and FTFT fabricated and measured MAPbI$_3$ samples for steady-state PL and solar cells. KPG measured SEM under the direction of ML. KPG measured XRD. KPG and ADT evaluated the time-dependent understoichiometric MAPbI$_3$ degradation. QA fabricated samples for PL mapping and fabricated and measured the MA-free samples. TS fabricated triple-cation solar cells under the supervision of ADT. YJH measured XPS. AY and AK measured the PL maps under the supervision of IS. KPG wrote the manuscript, and IS and YV edited. All authors reviewed the manuscript prior to submission.

## Conflicts of interest

There are no conflicts to declare.

## Acknowledgements

We thank Dr. Fabian Paulus for helpful discussions. Y.V. has received funding from the European Research Council (ERC) under the European Union's Horizon 2020 research and innovation programme (ERC grant agreement number 714067, ENERGYMAPS) and the Deutsche Forschungsgemeinschaft (DFG) in the framework of the Special Priority Program (SPP 2196) project PERFECT PVs (number 424216076). I. S. thanks the Swedish Research Council (2020-03530) and Knut and Alice Wallenberg foundation (2016.0059).

## Notes and references